\documentclass[preprint,prl,amsmath,amssymb]{revtex4}

\usepackage{graphicx}
\usepackage{dcolumn}
\usepackage{bm}

\def\kT{{k_{{}_B}T}}

\def\gsim{\mathrel{\raise.3ex\hbox{$>$\kern-.75em\lower1ex\hbox{$\sim$}}}}
\def\lsim{\mathrel{\raise.3ex\hbox{$<$\kern-.75em\lower1ex\hbox{$\sim$}}}}

\def\DA{{\Delta A}}

\def\DG{{\Delta G}}
\def\DFc{{\Delta G_{\rm channel}}}
\def\DFm{{\Delta G_{\rm memb}}}
\def\DFd{{\Delta G_{\rm dilate}}}
\def\DFt{{\Delta G_{\rm tilt}}}
\def\DFco{{\Delta G_{c_o}}}
\def\r{{\bf r}}
\begin{document}

\title{Gating-by-tilt of mechanosensitive membrane channels}
\author{Matthew S. Turner${}^{1}$ and Pierre Sens${}^2$}
\affiliation{${}^1$Department of Physics, University of Warwick, Coventry CV4 7AL, UK}
\affiliation{${}^2$Institut Curie, 11 rue Pierre et Marie Curie, 75231 Paris Cedex 05, France}

\date{\today}



\begin{abstract}
We propose an alternative mechanism for the gating of biological membrane channels in response to membrane tension that involves a change in the slope of the membrane near the channel. Under biological membrane tensions we show that the energy
difference between the closed (tilted) and open (untilted) states can far exceed $\kT$ and is comparable to what is available under simple dilational gating. Recent experiments demonstrate that membrane leaflet asymmetries (spontaneous
curvature) can strong effect the gating of some channels. Such a phenomenon would be more easy to explain under gating-by-tilt, given
its novel intrinsic sensitivity to such asymmetry.
\end{abstract}

\maketitle

\begin{figure}
\includegraphics{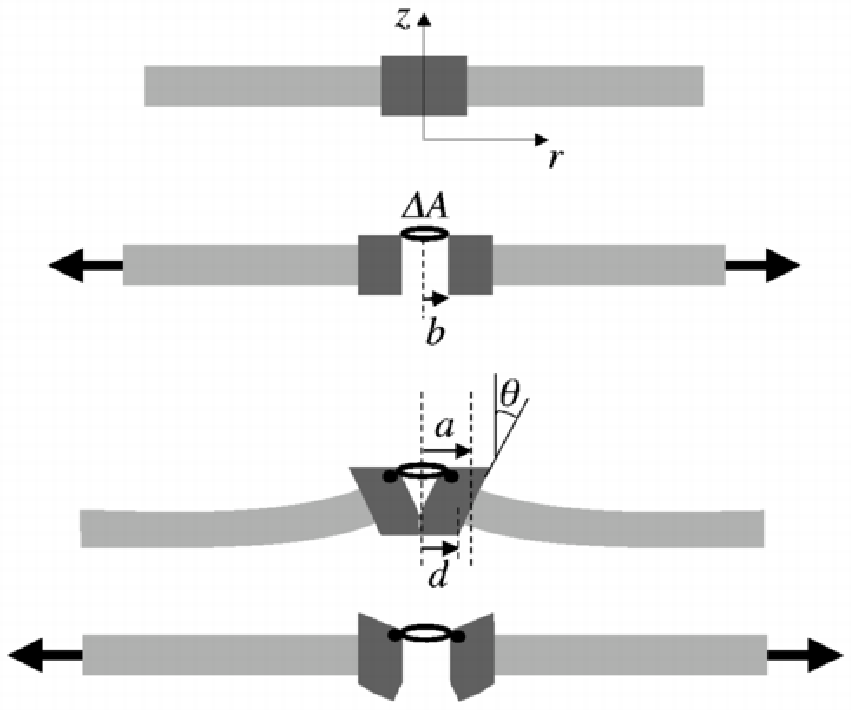}
\caption{Idealised sketch of a cylindrically symmetrical membrane channel showing two schemes for tension mediated gating. The upper two images show the closed and open states of the channel under the dilational gating model and the lower two
images the same for gating-by-tilt. Under both schemes the membrane tension does work by increasing the combined projected area of the membrane and channel.}
\label{gating}
\end{figure}

The correct biophysical function of mechanically sensitive membrane channels, such as the widely studied MscL\cite{sukharev93,sukharev94,sukharev99,sukharev01}, is vital in maintaining the viability of living cells. These channels are typically made up of a 5-8 transmembrane proteins that form a barrel-like assembly\cite{chang}. Fluid flows through a central pore in the open (active) state but is either restricted, or suppressed entirely, in the closed (inactive) state. As well as maintaining osmotic balance mechanosenstive channels play important sensing roles in touch, hearing, turgor control in plant cells etc \cite{sukharev97,hamill,booth,wood,martinac87,martinac01}. The fact that MscL functions in reconstituted membranes \cite{sukharev94} is good evidence for a gating mechanism that depends only on membrane tension, rather than cytoskeletal effects \cite{pickles} or signalling cascades.  We propose a new mechanism for the gating of biological membrane channels in response to elevated membrane tensions, which we will refer to throughout as gating-by-tilt, see Fig~\ref{gating}. The elevated membrane interfacial tension that acts to open the channel can be generated in several ways including via an osmotic imbalance between the cell interior and exterior or by changes in the cells morphology during adhesion, filipodia formation etc\cite{alberts}. 

The gating mechanism of such mechanosensitive channels have been probed by numerous patch clamp experiments (see e.g. \cite{hamill} and references therein) and have been studied fairly extensively by molecular dynamics (MD) simulations, see e.g. \cite{gullingsrud}. The
results of these studies have been interpreted as being consistent with gating that operates primarily via a process
resembling simple dilational opening of the channel, see Fig~\ref{gating}. However, both the experimental techniques and MD
simulations have limitations that make it difficult to be conclusive concerning this mechanism. Current computational limits
on the MD simulations allow only for only very short simulation times, typically of the order of a few nanoseconds, while it is known that gating takes milliseconds to occur after a step change in the patch clamp pressure that controls membrane tension\cite{hamill}. It is common practice in MD simulations to artificially apply thermodynamically large forces to the membrane (channel) in
order to induce it to open on the available timescales. It is therefore not clear that
these simulations are able to differentiate between certain models for how these channels gate. Indeed it is
one goal of the present work to propose a new paradigm for consideration in future simulations. Patch clamp experiments essentially measure the tension at which the channel opens and the size of its open pore by way of conductivity measurements\cite{hamill,kloda02,hille}. It isn't clear how these experiments can be used to differentiate between different gating models. Interestingly it is also known that the originally proposed pentameric structure of MscL \cite{chang} may be somewhat different from its {\it in vivo} structure \cite{hamill}. Thus it is possibile that some membrane tilt may be induced by the orientation of the walls of mechanosensitive channels in their closed configuration. Such a feature must be present if gating-by-tilt is to play a role.
During the opening of the channel the membrane tension does work to change the channel's conformation. The conformational free
energy of the channel itself (and alone) changes by an amount $\DFc\equiv G_{\rm open}-G_{\rm closed}$ which is
positive, indicating that the closed state has a lower conformational free energy. In what follows we will address only whether the channels are open or closed at equilibrium under constant membrane tension.
The existing paradigm for tension-gated channels involves a simple dilational transition from the closed to the open state with
an associated increase in total channel area $\Delta A$ \cite{hamill}. This is the increase in the total effective ``footprint"
of the channel within the membrane. The change in free energy is proportional to this and to the membrane tension $\gamma$ \begin{equation}\DFd=\gamma\Delta A \label{DFd}
\end{equation}
and is the only energy that is available to overcome the change in conformational free energy of the channel $\DFc$.
The proposed gating-by-tilt is driven by changes in the slope of the membrane where it meets the channel. The simplest
version of this involves a constant membrane slope $\theta$ in the closed state which relaxes ($\theta\to0$) after opening, see Fig~\ref{gating}. A central pore could be opened in association with such a transition. In practice a hybrid mechanism may operate in which there is both some change in tilt and some dilation. Other workers have attempted to study how changes in the lateral pressure profile of the channel might also affect gating\cite{perozo,cantor97,cantor99} although, in what follows, we will restrict our attention to effects that depend directly on the membrane tension.

We argue that it is {\it possible} for the action of the membrane tension $\gamma$ to efficiently gate the channel provided the
work done by the membrane $\DFm$ is large enough
\begin{equation}
\DFm(\gamma)\gsim\DFc\gsim\kT\label{DFmcriterion}
\end{equation}
where $\DFm=\DFd$ or $\DFt$, or a combination of the two. The criteria of Eq~(\ref{DFmcriterion}) equate to a channel
that {\it can} be substantially closed below tension $\gamma$ and substantially open above it. It is reasonable to assume that
nature has been able to evolve a channel with a configurational energy change that is roughly optimal for gating at the desired
tension, in which case one expects
$\DFm\approx\DFc$. If $\DFm/\kT$ is larger than unity then a substantial channel configurational energy can be overcome by the
action of the membrane tension and hence the channel can remain substantially closed at low tensions and substantially open at
high tensions (above the gating threshold). Alternatively if $\DFm/\kT$ is smaller than
unity the energy available from the membrane tension is inadequate to overcome any activation barrier that is greater than the
thermal energy scale and so the channel is either predominantly closed, if $\DFc>\kT$, or opens and closes rather randomly, if
$\DFc<\kT$. From Eq~(\ref{DFmcriterion}), and the arguments above, it is possible for a channel that opens by gating-by-tilt to efficiently
gate the channel at tension $\gamma$ when the parameter controlling this efficiency $\DFt(\gamma)/\kT\gsim 1$. We now proceed to
calculate the energy $\DFt$.

\begin{figure}

\centerline{\includegraphics[width=12cm]{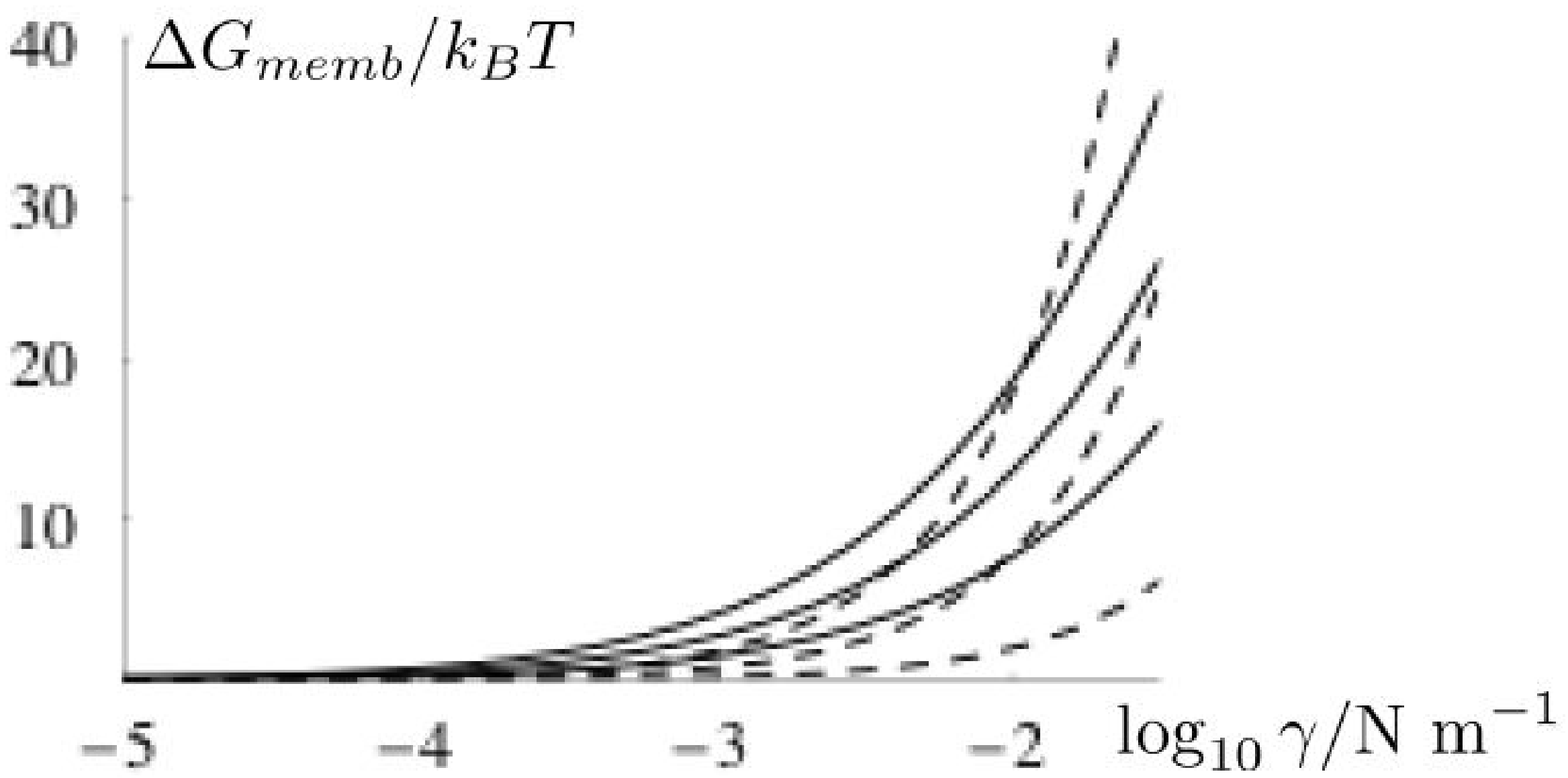}}
\caption{The work available from the coupling of the membrane tension to a conformational change of a membrane
channel $\DFm$ is plotted against the logarithm of the membrane tension $\gamma$. The dashed lines corresponding to the energy change $\DFd$ available from dilational opening (\ref{DFd}) and the solid lines the energy $\DFt$ available from gating-by-tilt (\ref{DFt}). Three curves are shown in each case corresponding to different values of the channel size. For dilational gating (dashed curves) the radius of the central pore in the open configuration is, from upper curve to lower curve, $b=1.5,\>1,\>0.5$ nm, while for gating-by-tilt the radial distance of the exterior of the closed channel is, from upper curve to lower curve, $a=4,\>3,\>2$nm. These represent biologically reasonable sizes \cite{kloda02,sukharev99}. In this illustrative example we have taken the change in tilt to be $\theta=\pi/6=30^{\circ}$ and the membrane rigidity $\kappa=20\kT$. The viability of the gating process relies on the value of the parameter $\DFm/\kT$ being larger than unity, as discussed in the text. Steady state biological membrane tension are in the $10^{-5}$N/m to $10^{-4}$N/m range \cite{sheetz} while Msc channels can gate at tensions up to $\gsim 10^{-2}$N/m \cite{kloda02}.}\label{DF}
\end{figure}

The Hamiltonian for membrane that is asymptotically flat, with normal parallel to the $z$-axis, but which has a small normal
deviation from flatness, of magnitude $u(\r)$ due, e.g. to the presence of a membrane channel, is given by \cite{safran}
\begin{equation}
H={1\over 2}\int d^2\r\left[\kappa(\nabla^2 u)^2+\gamma (\nabla u)^2\right]\label{H}
\end{equation}
where $\kappa$ is the membrane rigidity, $\r$ is the radial position with $r=|\r|$ (see Fig~\ref{gating}) and $\nabla$ is the two dimensional (plane polar) version of the operator. The total energy associated with distortion of the membrane can be established by a
variational approach on Eq~(\ref{H}) and depends on the boundary conditions for the membrane. Up to unimportant global rotations
or translations of the entire frame the displacement of membranes which are asymptotically flat at infinity is found
to be
\begin{equation}
u=\alpha K_0(kr)+\beta\log kr\label{ugeneral}
\end{equation}
for $r\ge a$ where $\alpha$ and $\beta$ are yet undetermined constants, $k\equiv\sqrt{\gamma/\kappa}$ is an inverse length
characteristic of the membrane and $K_0$ is the usual modified Bessel function of the first kind of order zero. It can be shown
that solutions of the form of $u\sim\log kr$ correspond only to channels which exert a finite integrated normal force on the
membrane\footnote{For further details see \cite{usPRE} but the analysis is somewhat analogous to electrostatics in 2D where a potential $\phi\sim\log r$ only appears if there is a finite total {\it charge} (here {\it force}) near the origin.}, e.g. by anchoring onto the cytoskeleton or an external substrate. Thus for channels which exert no
overall normal force we have $u=\alpha K_0(kr)$ where the constant $\alpha$ is fixed by a boundary condition corresponding to the
angular tilt at the periphery of the channel $\nabla u(r=a)=-\alpha k\>K_1(ka)=-\theta$. Hence
\begin{equation}
u=\theta K_0(kr)/(k\>K_1(ka))\qquad {\rm for}\, r\ge a\label{utheta}
\end{equation}
where the resulting normal membrane deviation (a few nm or less) falls to zero over a distance $\sim k^{-1}$ from the edge of the channel.The free energy difference $\DFt$ between a channel in the closed state and one in the open state
($\theta\to 0$, say) follows by substitution of Eq~(\ref{utheta}) into Eq~(\ref{H}) integrated over $r>a$. \begin{equation}
\DFt=\pi\kappa k a\theta^{2}{{K_0(ka)}\over{K_1(ka)}}\label{DFt}
\end{equation}
A quantitative comparison with $\DFd$ is shown in Fig~\ref{DF}.

For steady state physiological tensions in the range $\gamma=10^{-5}$--$10^{-4}$N/m one has $0.03< ka <0.1$ and the system is quantitatively inside the regime $ka\ll 1$ in which $\DFt$ can be shown to
have the following analytic approximation
\begin{equation}
\DFt\approx\pi\gamma a^2\theta^2\left[\log {2\over{ka}}-{\rm Euler\,Gamma}\right]\quad{\rm for}\,ka\ll 1\label{DFtasymptotic}
\end{equation}Where Euler Gamma has a constant value of about $0.6$. At the highest gating tensions $ka$ can approach unity. Interestingly Eq~\ref{DFd} and Eq~\ref{DFtasymptotic} appear to be very similar, both being the product of an area change and the tension $\gamma$ except for the factor of $\log{2\over{ka}}$ appearing in Eq~(\ref{DFt}).
In order to test whether gating-by-tilt is indeed a plausible mechanism that may contribute to $\DFm$ we use data collected for various Mechanosenstive channels \cite{kloda02}, from which estimates for the the area change $\DA$, gating tension and conformational energy change $\DG$ have been obtained, assuming dilational gating. This gives us enough information to calculate the corresponding value for $\DFt$ from Eq~\ref{DFt}, assuming instead a purely gating-by-tilt scheme that occurs at the same tension. Since this energy will depend on the angle $\theta$ a sensible approach seems to be to ask instead what value of $\theta$ would be required to {\it entirely} account for the free energy change $\DG$ (previously estimated assuming dilational gating) and also what angle would be required to account for (say) 10\%\ of it. We will denote these angle $\theta_{100}$ and $\theta_{10}$ respectively. If a channel opens, via gating-by-tilt, to expose a pore of area $\DA=\pi b^{2}$ then we impose the boundary condition that the membrane slope is $\theta$ in the closed state at radial distance $r=a=d+b/2$ from the central ($z$) symmetry axis of the pore, where $d$ is the thickness of the ``walls'' of the channel, see Fig~\ref{gating}. This is a somewhat arbitrary, although reasonable, choice for $a$ which is merely the radial distance to the exterior of channel in the closed state. For the six channels for which data is available \cite{kloda02} we find (taking $d=2$nm) that $39^{\circ}>\theta_{100}>16^{\circ}$, while $12^{\circ}>\theta_{10}>5^{\circ}$ is correspondingly smaller. We find that the smallest values of these angles both correspond to the channel known as MscMJ which gates at the lowest value of $\DFt$ although {\it not} the lowest membrane tension. This may therefore represent the best candidate for future MD or experimental studies on channels that gate by tilt. The above angles are (all) rather small and thus we have demonstrated that a small change in the tilt of the membrane during gating can provide a significant contribution to the free energy change $\DFm$.

It has also been observed that the composition of lipids in the membrane can have a significant effect on the gating of MscL
channels \cite{perozo}. It is possible to construct arguments that can explain a dependence on membrane composition in terms of
mismatches between the membrane (thickness) and the channel geometry, or the pressure distribution exerted by the membrane
interior on the edge of the channel. However the dramatic effect recently observed on addition of ``conical"
lysophophatidylcholine (LPC) lipids to one leaflet of the membrane \cite{perozo} are generically difficult to explain within such
models. In these experiments it was observed that the channels will open, even under small applied tensions, only if the
conical lipids are localized in one leaflet, giving rise to an asymmetry that tends to make the membrane prefer to bend into a
convex shape, away from the LPC rich face, rather than the opposite concave one. It is possible to analyze this difference
within a model that has the LPC homogeneously distributed across the membrane or, experimentally, the membrane patch that is
clamped. In this case the Hamiltonian for the membrane includes the so-called spontaneous curvature $c_o$, which is a function
of the difference in LPC concentrations across the inner and outer leaflets of the membrane. If the LPC is localised in the upper leaflet $c_o<0$.  When the spontaneous curvature is small $c_{o}^{2}\ll k^{2}$ one obtains
\cite{safran}
\begin{equation}
H={1\over 2}\int d^2\r\left[\kappa(\nabla^2 u-c_o)^2+\gamma (\nabla u)^2\right]\label{Hco}
\end{equation}
The corresponding minimum energy solution Eq~(\ref{utheta}) for the
displacement is unchanged under these conditions. However the total free energy of the membrane can be shown to be $\DFm=\DFt+\DFco$ with $\DFt$ as before (\ref{DFt}) and
\begin{equation}
\DFco=2\pi c_o\kappa\theta a\label{DFco}
\end{equation}
This free energy difference between the open and closed state can be positive, favoring opening, or negative, helping to
maintain a closed channel if $\theta>0$ or $\theta<0$ respectively. The sign of $\theta$ depends on whether the the channel is
oriented either ``up" or ``down". In experiments on reconstituted membranes it is very likely that the channels will be in both
orientations. Eq~(\ref{DFco}) suggests that for moderate spontaneous curvatures $c_o=1/(10a)$ the ``up'' channels (say) are driven to open by an energy difference $\DFco\approx 7\kT$, even in the absence of any tension. This gives a Boltzmann weight of about 1000, with a corresponding enhancement in the number of open channels. This could help to explain why some
channels isolated in a patch clamped membrane are observed to remain open under these conditions, even under small
tensions (pressure differences) \cite{perozo}. Indeed these authors independently suggest that, ``The asymmetry of the lateral pressure profile between the two leaflets of the bilayer is what actually initiates the sequence of mechanical transduction steps that leads to the open state". We argue that such effects are highly suggestive of a role of the gating-by-tilt mechanism proposed here.
In conclusion the energy difference available between between the closed (tilted) and open (untilted) channel conformations is
found to be comparable to that for dilational opening, as estimated from experimental measurements, and could be even greater for
artificially engineered membrane systems. Thus we propose that the gating-by-tilt mechanism should be considered a viable candidate for channel gating, particularly in view of the lack of precise knowledge concerning the channel geometries in the open and closed states. There seems no obvious {\it a priori} reason why gating-by-tilt should not be commonly employed in nature, short of some generic difficulty in bio-engineering a channel architecture that gates-by-tilt. Furthermore, under gating by tilt, the tilting walls of the channel can be attached by a hinge to a fairly rigid frame, from which they swing open. This suggests that gating-by-tilt may have further inherent design advantages over dilational gating in which the {\it entire} channel must move (dilate).  Since preparing this work we have learned of an independent study \cite{caltech} that investigates the gating of mechanosensitive channels via effective line tensions. These would also appear within our model for large $ka\gg 1$, although we make no assumptions about the value of this parameter, which rarely seems to greatly exceed unity. Furthermore, these authors seem to assume that any channel tilt remains fixed throughout channel opening, in which case the tilt uniformly favors the closed (undilated) state. This is fundamentally different to the present work in which we propose tilt {\it variation} as the gating mechanism.

\newpage

\newpage

\end{document}